\documentclass[runningheads]{cl2emult}

\usepackage{makeidx}  % allows index generation
\usepackage{graphicx} % standard LaTeX graphics tool
\usepackage{psfig}
\usepackage{subeqnar} % subnumbers individual equations
\usepackage{multicol} % used for the two-column index
\usepackage{cropmark} % cropmarks for pages without
\usepackage{eso}      % placeholder for figures
\makeindex            % used for the subject index
\newcommand{\ltapprox}{\raisebox{-0.5ex}{$\,\stackrel{<}{\scriptstyle\sim}\,$}}
\newcommand{\gtapprox}{\raisebox{-0.5ex}{$\,\stackrel{>}{\scriptstyle\sim}\,$}}

\begin{document}
\title*{Probing distant galaxies with lensing clusters}
\toctitle{Probing distant galaxies with lensing clusters}
% allows explicit linebreak for the table of content
%
%
\titlerunning{Probing distant galaxies with lensing clusters}
% allows abbreviation of title, if the full title is too long
% to fit in the running head
%
\author{Roser Pell\'o\inst{0}}

\authorrunning{R. Pell\'o}
% if there are more than two authors,
% please abbreviate author list for running head
%
%
\institute{Observatoire Midi-Pyr\'{e}n\'{e}es,
           LAT, UMR5572,
           14, Av. Edouard-Belin,\\
           F-31400 Toulouse, France}

%\and Universit\'{e} de Paris-Sud,
%     Laboratoire d'Analyse Num\'{e}rique,
%     B\^{a}timent 425,\\
%     F-91405 Orsay Cedex, France}

\maketitle              % typesets the title of the contribution

\footnotetext{
In collaboration with :
%the Gravitational Telescope team: 
B. Fort (IAP, F),
J.-P. Kneib (OMP, F),
J.-F. Le Borgne (OMP, F),
Y. Mellier (IAP, F),
I. Appenzeller (Obs. Heidelberg, D),
R. Bender (Munich Observatory, D),
L. Campusano (U. Chile, Chile),
M. Dantel-Fort (O. Paris, F),
R.S. Ellis (IoA, UK)
A. Moorwood (ESO, D),
S. Seitz (Munich Observatory, D)
}

\begin{abstract}

   Clusters of galaxies as gravitational lenses allow to study the
stellar content and properties of high-z galaxies much fainter than
the usual spectroscopic field surveys. We review the recent results
obtained on the identification and study of very distant galaxies
seen through lensing clusters. Using the gravitational amplification
effect in lenses with well known mass distributions, it is possible 
to build up and study a sample of galaxies with $1 \ltapprox z
\ltapprox 7$. Source candidates are selected close to the critical 
lines at high-z, through photometric redshifts computed on a large 
wavelength interval, as well as lens-inversion criteria. This procedure
allows to reduce the selection biases in luminosity, and towards active
star-forming objects, provided that selection criteria include
IR photometry. This is the method presently used in our large collaborative 
program, aimimg to perform the spectroscopic follow up with the VLT
of high-z candidates selected from suitable photometric campaigns.

\end{abstract}

\section{The Gravitational Telescope}

The identification and study of high-z galaxies is probably one of the most 
direct methods to constrain the scenarios of galaxy formation and evolution.
With respect to large field surveys, 
cluster lenses can be used to build up an {\it independent} and complementary
sample of distant galaxies, because of their lensing properties. The major bonus 
is the amplification close to the critical lines, which is $\Delta m \sim $ 1 
to 3 magnitudes depending on location (see \cite{Fort} for a review). 
The expected 2D distribution of arclets (mean redshift and surface density) 
can be easily obtained for lenses with well known mass-distributions 
(\cite{Beze}). These clusters are used here as Gravitational Telescopes . 

The first lensed galaxy confirmed at $z \gtapprox 2$ was the spectacular
blue arc in Cl2244-02 (\cite{Mellier}). Recent
examples of highly magnified galaxies, identified either purposely or
serendipitously, strongly encourages this approach: the star-forming source
$\#384$ in A2218, at z=2.51 (\cite{Ebbels1}), the luminous z=2.7 galaxy 
behind MS1512+36 (\cite{Yee}), three z
$\sim$ 4 galaxies in Cl0939+47 (\cite{Trager}), a z=4.92 system in
Cl1358+62 (\cite{Franx}, \cite{Soifer}), the z=2.72 arc behind 
MS1512+36 (\cite{Seitz}), and the two red galaxies at
$z \sim 4$ in A2390 (\cite{Frye}, \cite{Pello}).
We present the method proposed by the Gravitational Telescope Program
(hereafter, GT) to access the distant 
population of galaxies and some illustrative results.

\begin{figure}
\centering
\noindent
\centerline{\hbox{
\includegraphics[width=.5\textwidth]{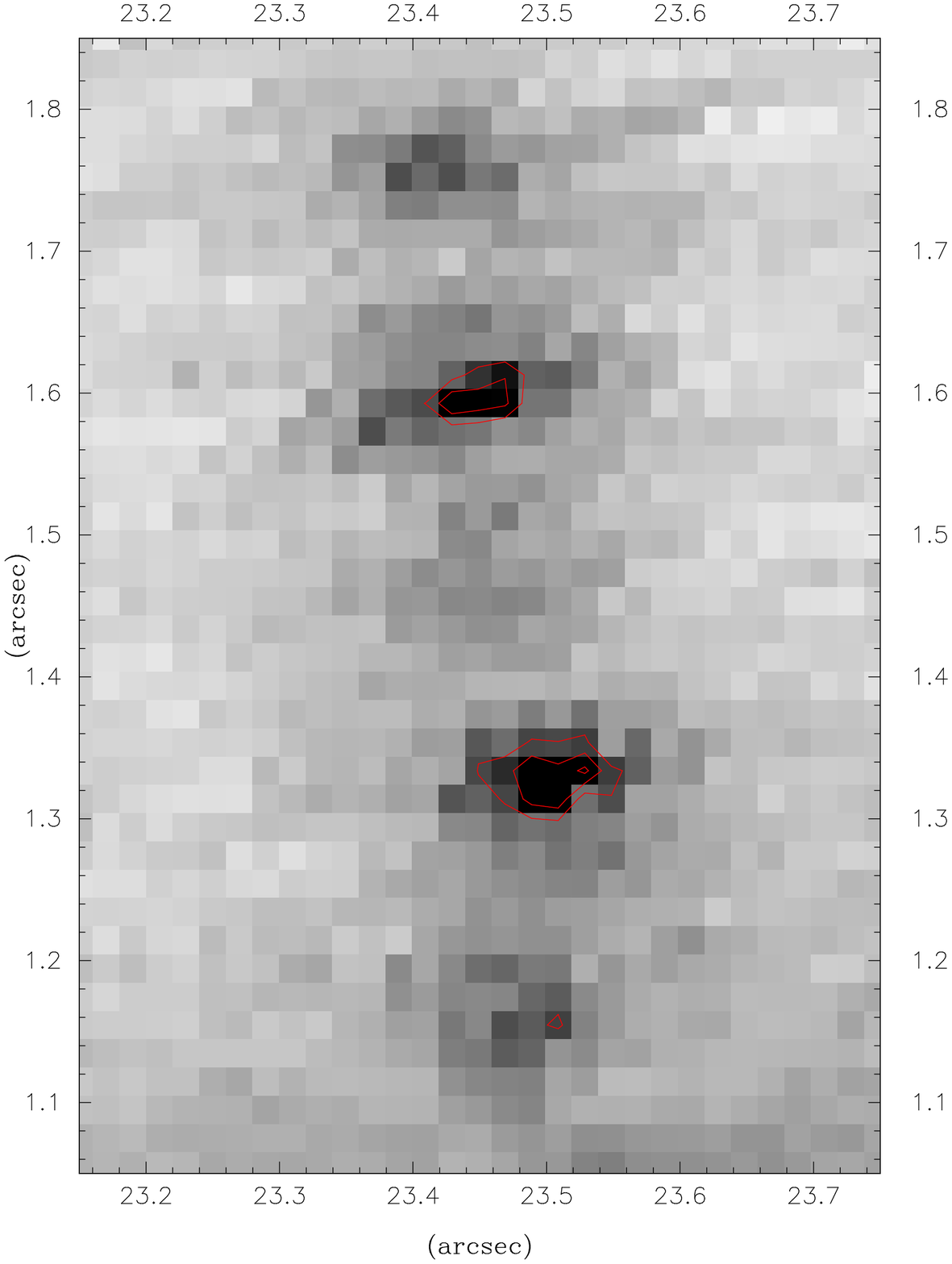}
\includegraphics[width=.5\textwidth]{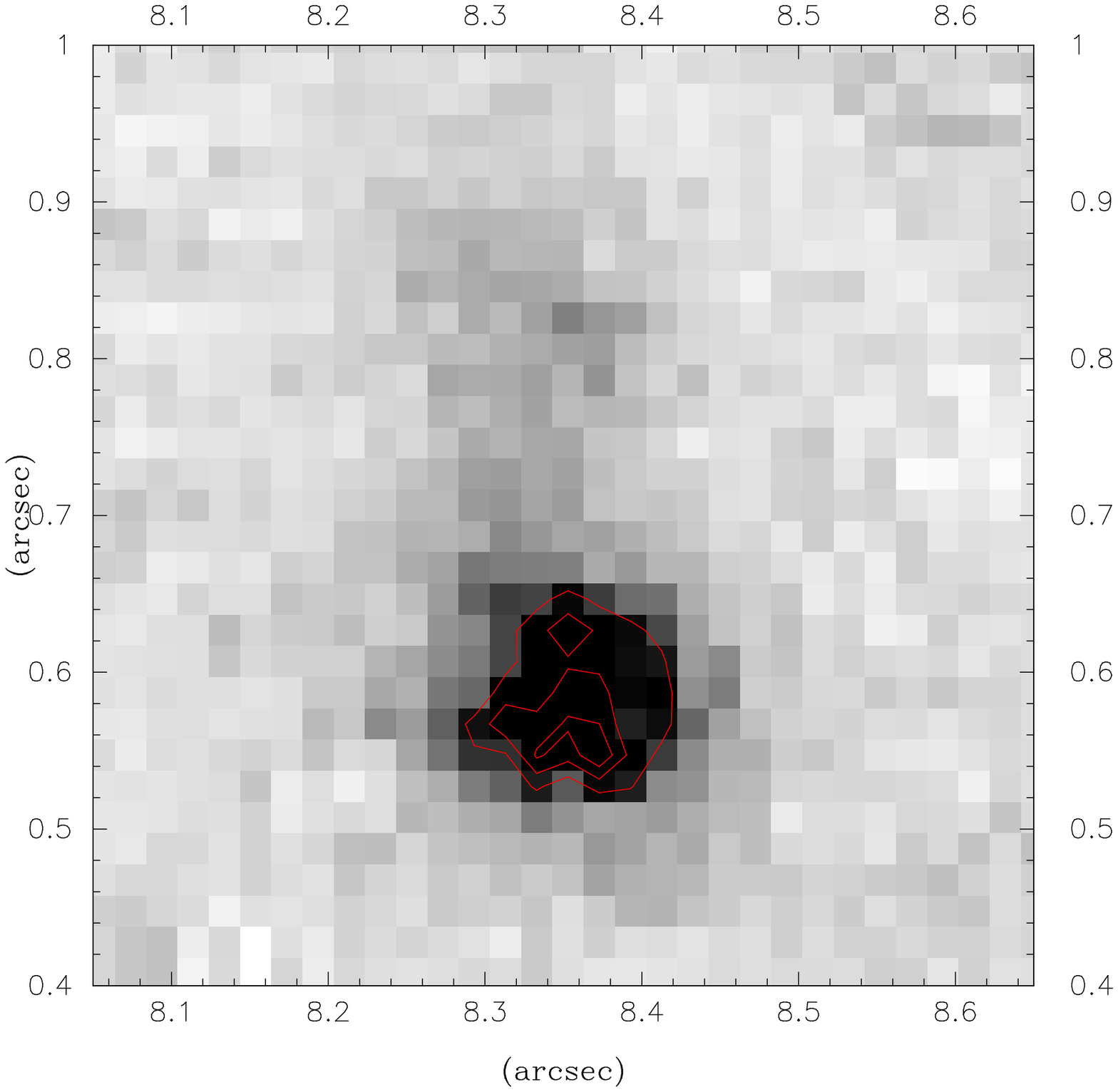}
}}
\caption{Images of 2 $z = 4.05$ sources behind A2390:
H3 ({\bf left}) and H5 ({\bf right}) (\cite{Pello}), 
reconstructed by lens-inversion 
%on their source plane,
according to the model by Kneib et al. (1999).
Both sources are clumpy, elongated and exhibit the same orientation. 
The true distance between H3 and H5 is $16''$
($1'' = $ $9.4 (15.7) h_{50}^{-1}$ kpc with $q_0= 0.5 (0.1)$)}
\label{fig1}
\end{figure}

\section{The Method}

Lensed sources in the GT program are selected according to 2
criteria:  {\bf 1)} they lie close to the high-z critical lines,
and {\bf 2)} they have a photometric redshift compatible with $z \ge
2$. The former criterion restricts the application to clusters with
mass distributions highly constrained by multiple images (revealed by
HST or ground multicolor images), where the amplification 
uncertainties are $\Delta m_{lensing}<$ 0.3 mags.
Cluster lenses with well constrained mass distributions enable to
recover precisely the properties of lensed galaxies (morphology, 
magnification factor). Fig. 1 gives an example of source reconstruction
for two multiple-images at the same $z \sim 4$, behind
A2390. The sample of lenses includes clusters with a
strong X-ray emission and, as a first priority, those with HST images available. 
In all cases, deep multicolor photometry has been obtained in the
visible and near-IR domains.

Photometric redshifts have been derived using the standard 
SED fitting method by Miralles \& Pell\'o (1998). 
The observed SEDs are compared to a set of 
templates linked to the Bruzual \& Charlot code (GISSEL98,
\cite{Bruzual}). The database includes 255 synthetic spectra,
with $A_v$ from 0 to 0.5 mags, and flux decrements in the 
Lyman forest modelled 
according to Giallongo \& Cristiani (1990; see also Madau 1995). 
The photometric z criterion can be used advantageously in this case
because photometry is available on a large wavelength domain, including
near-IR. This selection allows to reduce the biases 
towards intrisically bright or strong star-forming sources. Fig. 2 presents 
an example of likelihood map for one of the $z \sim 4$ sources in Fig. 1. 

\section{Main goals}

The GT can be used to determine the redshift 
distribution up to the faintest levels through magnified sources.
It is also the natural way to search for primeval galaxies, in
order to put strong constraints on the scenarios of galaxy formation.
The use of visible and near-IR spectroscopy, combined with 
photometric z, enables to study the SED of $z \ge 2$ 
galaxies for a sample which is less biased in luminosity than the field ones.
In particular, the present and past SFR history (respectively obtained 
through the UV/visible/IR and near-IR domains),
and the permitted region in the age- metallicity -$A_V$
parameter space.
Highly magnified arcs are presently the only way to access the dynamical 
properties of galaxies at $z \ge 2$, through 2D spectroscopy
(Narasimha \& Chitre, 1993; Soucail et al. 1998). As shown in Fig.1,
lens reconstruction allows to explore the morphology of distant 
sources with a spatial resolution of $\sim$ 1 kpc. When this technique
will be coupled with 2D spectroscopy, the dynamics involved in
galaxy formation would be directly observed.

\section{First results on selected lenses}

H3 and H5 are the first two $z \ge 2 $ objects spectroscopically confirmed
in A2390, and this is one of the best studied lenses in our sample. 
$\sim 30$ sources at $z \ge 2 $  have been identified among the numerous arcs and
arclets in this field.
Their redshift distribution is shown in Fig. 5, and compared to the 
present spectroscopic survey in A2390. The redshift distribution in 
other GT targets (A2218, A370, Cl2244-02, AC114) also displays interesting
candidates at $1 \ltapprox z\ltapprox 7$.
Once spectroscopic data is available for high-z sources, the
SF rates and other stellar properties can be derived, although there
is some degeneracy in the SFR-age-metallicity-reddening space.
The permitted parameter region can be constrained using the GISSEL98 code
when z is well known.
Fig. 4 illustrates this point with two examples : H5 and D, the 
later being one of the brightest ISO sources in A2390 (\cite{Lemon}, 
\cite{Altieri}).
H3 and H5 are intrinsically bright star-forming systems
($M^{*}_B -2$ to -1 mags, depending on
$A_v$), slightly brighter than the $z \sim 4$ objects found 
in Cl0939+4713 \cite{Trager}.
%, and also brighter than the $z=5.34$
%galaxy of Dey et al. (1998).

For a subsample of spectroscopically confirmed objects, we have tested the 
photometric z accuracy as a function of the SFR,
reddening, age and 
metallicity of the stellar population. Spectroscopic
surveys were carried out at CFHT, WHT and ESO (NTT,3.6m). We have cross-checked 
the consistency between the photometric, the spectroscopic and the lensing z
obtained from inversion methods (\cite{Ebbels2}). The agreement
between the three methods is good up to at least $z \ltapprox 1.5$. For
higher redshifts, the results on the most amplified 

\begin{figure}
\centering
\noindent
\centerline{\vbox{
\includegraphics[width=.7\textwidth,angle=270]{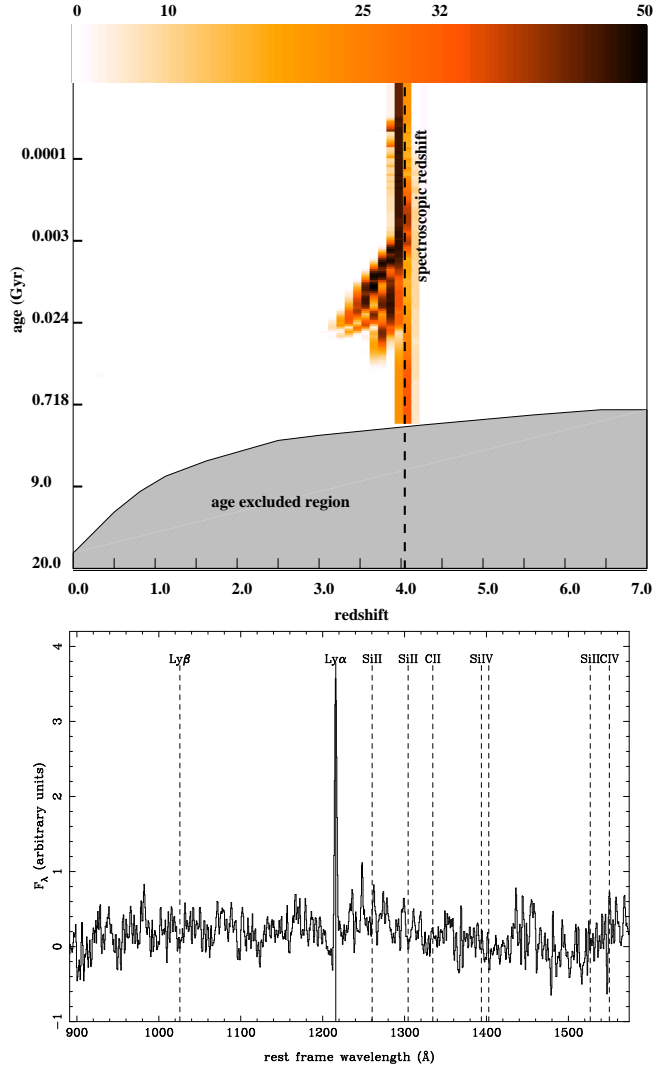}
\includegraphics[width=.5\textwidth,angle=270]{pello_f2b.ps}
}}
\caption{H5, one of the z=4.05 sources in A2390: 
{\bf Top) } Photometric z likelihood-map showing the 
excellent agreement with the spectroscopic z.
The shaded region is the limit set for
the age of the Universe ($H_0$ = 50 km s$^{-1}$ Mpc$^{-1}$,
$q_0=0$).
{\bf Bottom) } Averaged spectrum of H5, obtained at WHT,
showing a strong Ly$\alpha$ emission line
}
\label{fig2}
\end{figure}

\noindent sources are
promising, but an enlarged spectroscopic sample is needed.

\section{Future issues}

%\begin{minipage}{4.0cm}
\begin{figure}
\centering
\includegraphics[width=.5\textwidth,angle=270]{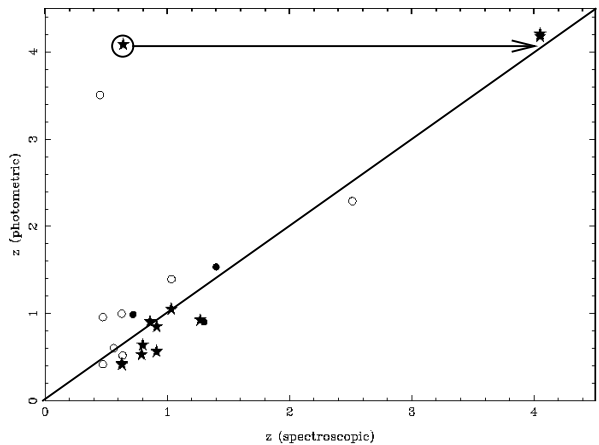}
\caption {Spectroscopic versus photometric redshift determination 
%{\bf Fig. 3.} Spectroscopic versus photometric redshift determination 
for a sample of multiple images in A370 (black dots), A2390 (black stars)
and A2218 (open dots). One of the two catastrophic identifications corresponds
to a wrong spectroscopic determination based on a single emission line}
\label{fig3}
\end{figure}
%\end{minipage}

\begin{figure}
\centering
\noindent
\hbox{
\includegraphics[width=.5\textwidth]{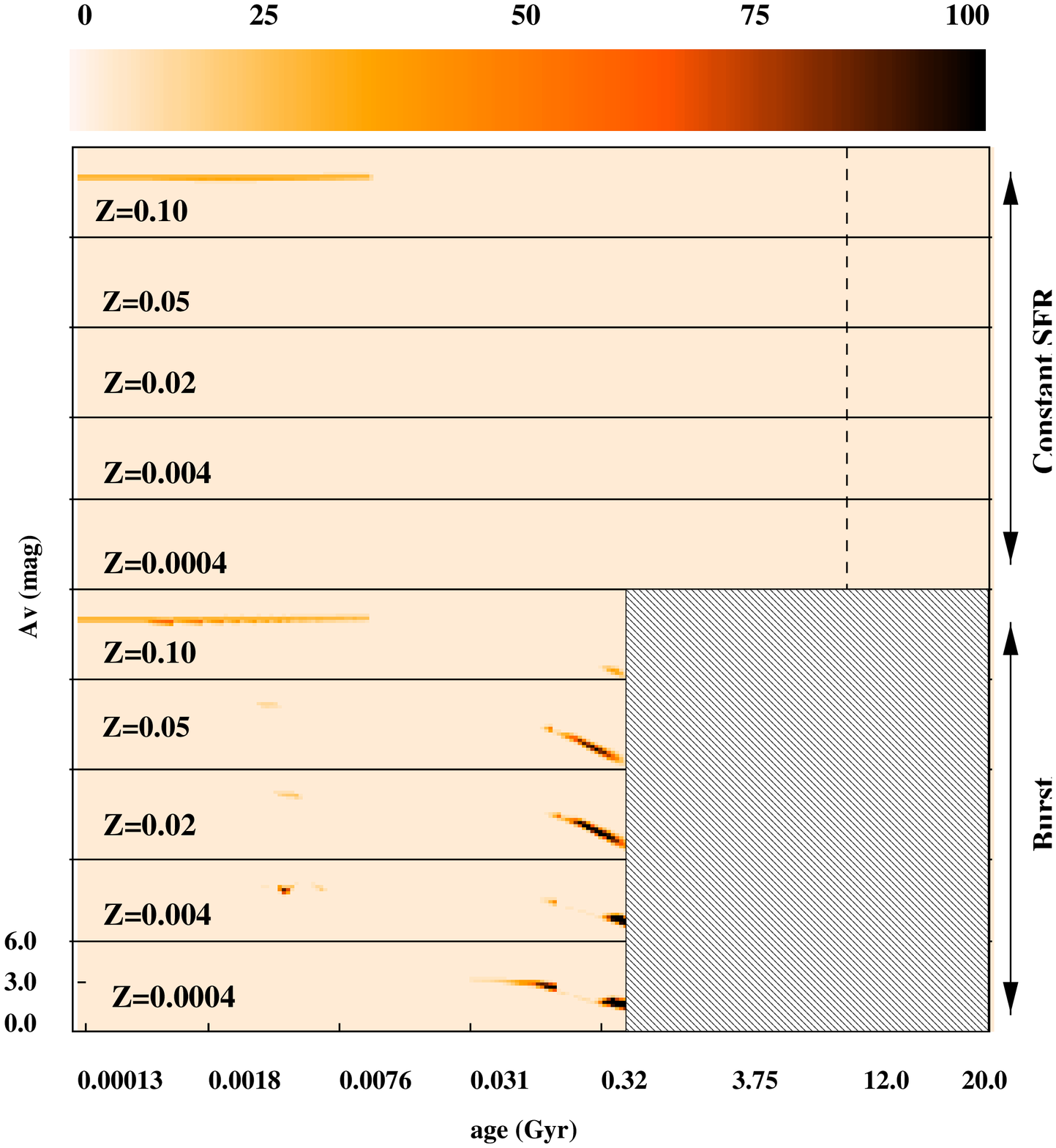}
\includegraphics[width=.5\textwidth]{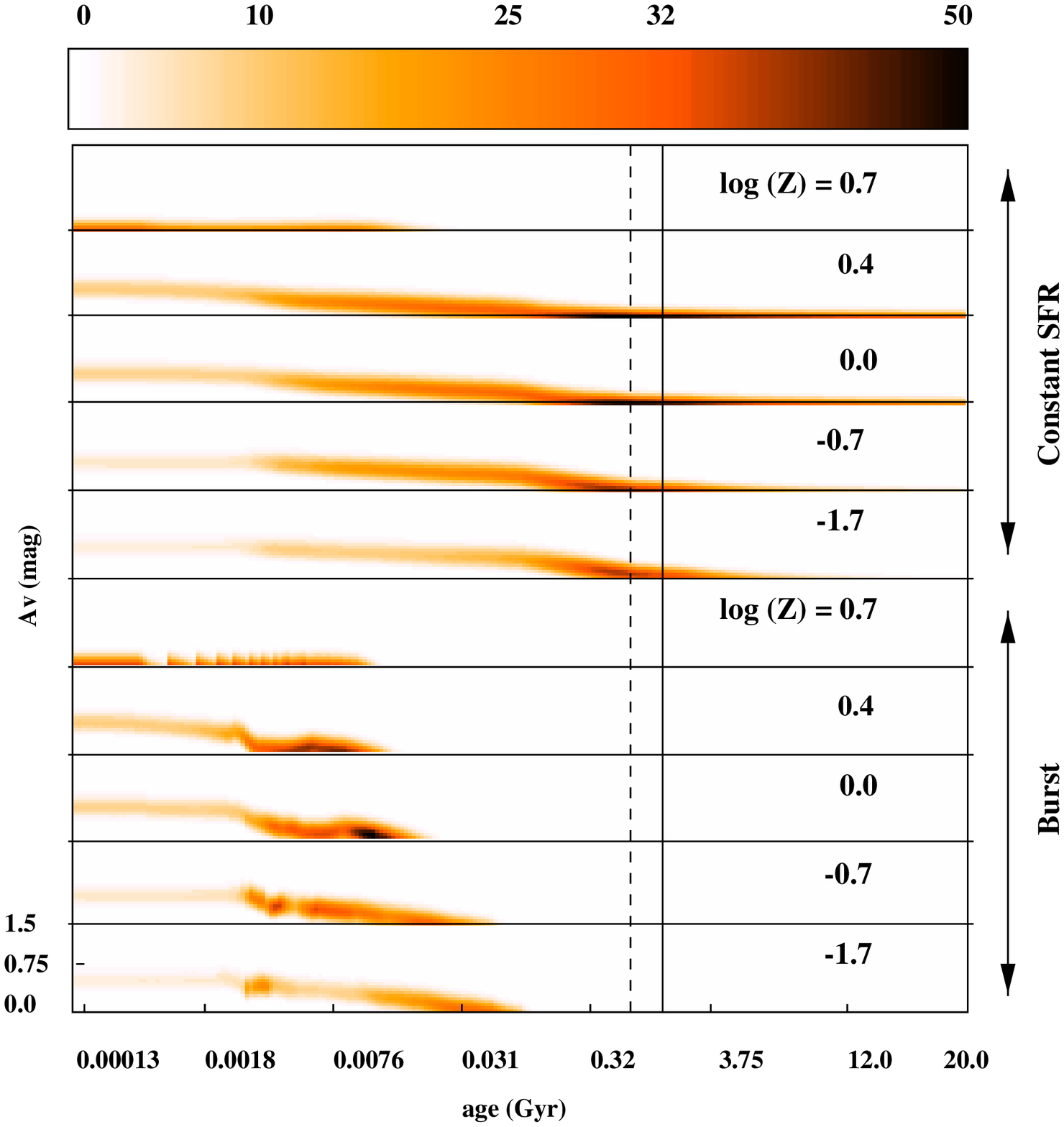}
}
\caption
{Likelihood map for two sources in A2390, showing in dark the most probable
regions and the degeneracy in the parameter space SFR - age - metallicity -
reddening: {\bf left) } Source D (z=0.913): the dotted line gives the age limit of 
the stellar population ($q_0=0$); the hatched region is 
forbidden according to the optical spectrum, which shows a strong $[OII]3727 \AA$ \
e-line, and evidence for ongoing star-formation. The most likely solutions 
exhibit high reddening, as expected for a such bright ISO source.
{\bf right) } H5 at z=4.05; dotted and solid lines are the cosmic ages
($H_0$ = 50 km $s^{-1} Mpc^{-1}$, $q_0=0.1$ and $q_0=0.5$)}
\label{fig4}
\end{figure}

\begin{figure}
\centering
\noindent
\includegraphics[width=0.65\textwidth,angle=270]{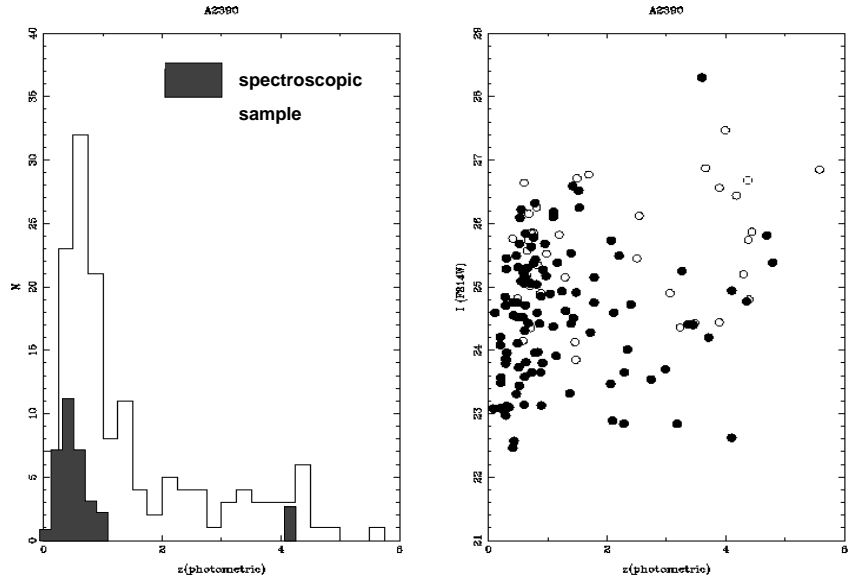}
\caption{ {\bf left) } Photometric redshift distribution for 
arclets in A2390, for the most secured sources (black dots). 
{\bf right)} I (F814W) magnitude versus photometric z
}
\label{fig5}
\end{figure}

The uncertainty in the amplification is typically $\Delta m \sim $ 0.2
to 0.3 mags. Thus, intrinsic luminosities and SFRs are known 
to $\sim 30\%$ accuracy using the GT. Only well
studied lensing clusters, with fairly well known mass distributions, are 
actually useful as GTs.

Up to now, all spectroscopically confirmed sources at high-z in
our sample are
intrinsically bright. In order to constrain their parameter space 
(metallicity, $A_V$, ...), spectroscopic information on the 
UV restframe absorption lines, and near-IR spectroscopy aiming to access
the Balmer region are strongly needed. Using arcs in clusters is
probably the only way to access dynamics for $z \ge 2$ sources
(velocity gradiengts, 2D velocity fields, ...). This systematic 
study is a well defined program for 8m telescopes such as the VLT.
Combining photometric redshifts with lensing inversion techniques
provides with an alternative way to determine the redshift distribution 
of high-z sources. If $\Delta z \sim 0.1$ is enough for most applications, 
photometric redshifts allow to go further. Lensing clusters could be 
used as a tool to check photometric redshifts up to the faintest
limits. An Ultra-Deep Photometric Survey of cluster lenses is urgently needed
to probe the distant Universe.

%INDEX%%%%%%%%%%%%%%%%%%%%%%%%%%%%%%%%%%%%%%%%%%%%%%%%%%%%%%%%%%%%%%%
\clearpage
\addcontentsline{toc}{section}{Index}
\flushbottom
\printindex
%%%%%%%%%%%%%%%%%%%%%%%%%%%%%%%%%%%%%%%%%%%%%%%%%%%%%%%%%%%%%%%%%%%%%

\end{document}